\documentclass[prd,preprint]{revtex4}
\newcommand{\ch}[1]{{\color{black} #1}}
\usepackage{url}
	\newcommand{\hl}[1]{{\color{black} #1}}
	\usepackage{amsmath}%,amssymb} 
	\usepackage{amsfonts}
  	\usepackage{amssymb}
  	\usepackage{float}
	\usepackage{makeidx}
	\usepackage{amsfonts}
	\usepackage[ansinew]{}
	\usepackage[usenames,dvipsnames]{pstricks}
	\usepackage{booktabs}
	\usepackage{multirow}
	\usepackage{epsfig}
\usepackage{xcolor} 

\newcommand{\be}{\begin{equation}}
\newcommand{\ee}{\end{equation}}
\newcommand{\bea}{\begin{eqnarray}}
\newcommand{\eea}{\end{eqnarray}}

\makeindex

\begin{document}

%\preprint{APS/123-QED}

\title{Combining gravitational and electromagnetic waves observations to investigate local structure and the Hubble tension}% Force line breaks with \\

\author{Brayan Yamid Del Valle Mazo${}^{2,3}$}
\author{Antonio Enea Romano${}^{1,2,3}$}
\author{Maryi Alejandra Carvajal Quintero${}^{2,3}$}

\affiliation{%
${}^{1}$Theoretical Physics Department, CERN,CH-1211 Geneva 23, Switzerland\\
${}^{2}$ICRANet, Piazza della Repubblica 10,I–65122 Pescara, Italy\\
${}^{3}$Instituto de Fisica,Universidad de Antioquia,A.A.1226, Medellin, Colombia
}%

\date{\today}% It is always \today, today,

\begin{abstract}
Recent estimations of the Hubble parameter $H_0$ based on gravitational waves (GW) observations can be used to shed some light on the discrepancy between the value of the Hubble parameter $H_0^P$ obtained from large scale observations such as the Planck mission, and the small scale value $H_0^R$, obtained from low redshift supernovae (SNe).  

In order to investigate the origin of this discrepancy we perform a combined analysis of the  luminosity distance of SNe and GW sources, using different methods, finding that the impact of the GW data is very limited, due to the small number of data points, and their large errors.  We analyze separately data from the Pantheon and the Union 2.1 catalogues, finding that a model with $H_0^P$ and a small local void can fit the data as well as a homogeneous model with $H_0^R$, resolving the apparent $H_0$ tension. We find that there is a significant difference between the size and depth of the inhomogeneity obtained using the two datasets, which could be due to the different sky coverage of the two catalogues.
For  Pantheon  we obtain evidence of a local inhomogeneity with a density contrast  $\delta_v=-0.155 \pm 0.026$, extending up to a redshift of $z_v = 0.056 \pm 0.0002$, while for Union 2.1 we obtain $\delta_v=-0.461 \pm 0.032$ and $z_v= 0.081\pm 0.008$.
We also perform some analysis using redshift shell averaged data, and obtain approximately the same results, hinting to the fact that the effects of the monopole component of the local inhomogeneity are the dominant ones. %Due to the limited number of GW sources, the results of the fits depend only mildly on the inclusion of GW data in the analysis. Observational data from future GW detectors and SNe surveys will allow to further investigate the presence of local structure and its implications on the estimation of cosmological parameters. 
%The monopole effect of this inhomogeneity on the luminosity distance is proportional to the volume average of the density contrast inside the sphere of radius equal to the distance of the source, which is suppressed outside the inhomogeneity by the inverse cube of the distance. As a direct consequence, since most of the GW sources are located outside this inhomogeneity,  they are only negligibly affected by it. 
\end{abstract}

\pacs{Valid PACS appear here}% PACS, the Physics and Astronomy
\maketitle

%\tableofcontents
\section{Introduction}
\hl{Recent gravitational wave observations  \cite{Abbott:2019yzh} have provided a new estimation  of the Hubble constant $H_0$, and could be used to shed some light on the tension between the  large scale estimations based on the cosmic microwave background (CMB) radiation \cite{Akrami:2018odb}, and the value obtained analyzing low redshift SNe \cite{Riess:2016jrr}.} The latter analysis is based on the assumption that the Universe is well described by a spatially homogeneous solution of the Einstein's equations, but only an unbiased analysis of observations can actually confirm this  hypothesis. \hl{As an alternative to dark energy it was proposed  the existence of a very large local void \cite{Celerier:1999hp,Enqvist:2006cg,GarciaBellido:2008nz}, but large void models were incompatible with other observations, such as CMB. It was then investigated the effect of inhomogeneities in presence of dark energy \cite{Romano:2011mx}, showing how they could lead to a correction of the apparent value of the cosmological constant, or affect  \cite{Romano:2016utn,Fleury:2016fda} the Hubble diagram. Other studies estimate the variance of $H_0$ using different analytical and numerical methods \cite{Odderskov:2014hqa,Ben-Dayan:2014swa,Marra:2013rba}, but they do not perform an actual analysis of the observational data, so are not directly comparable to our results. We fit data without any homogeneity assumption. %, correcting some errors in previous calculations \cite{Kenworthy:2019qwq}, which were inconsistently giving effects extending far beyond the edge of the inhomogeneity.
} 
Our analysis is partially confirming the existence of a local underdensity  surrounding us  in different directions \cite{Keenan:2013mfa,Chiang:2017yrq}. 

\hl{There have been different approaches to the explanation of the $H_0$ tension \cite{Vagnozzi:2019ezj,DiValentino:2019jae,Conley:2007ng}, and it has been recently \cite{Kenworthy:2019qwq} claimed that there is no evidence of a local inhomogeneity, but in that analysis a fixed value of the absolute magnitude M was used, ignoring how it should change if $H_0$ is different, and instead of fitting directly the apparent magnitude, the intercept $a_b$ was used, whose mathematical definition relies implicitly on some homogeneity assumptions. } 
Here we will not propose any modification of the standard cosmological model, but only perform an unbiased analysis of SNe and GW sources luminosity distance data, in search of a possible evidence of peculiar velocity fields normally ignored when assuming a Friedmann–Lemaître–Robertson–Walker (FLRW) metric.

While number count observations only allow to measure directly the baryonic matter distribution,  other effects such as gravitational lensing, allow to measure the total matter density, including the dark matter component, since gravitationally baryonic and dark matter produce the same gravitational effects.
One possibility to overcome the difficulty of deducing the total density field from number counts, due for example to selection effects, is to  reconstruct the total matter density  distribution from the effects it imprints on the luminosity distance of  standard candles \cite{Chiang:2017yrq}, and standard sirens \cite{Abbott:2017xzu}. % i.e. astrophysical objects whose absolute magnitude is supposed to be approximately the same, such as for example Supernovae (SN) Ia or Cepheids.
 At low redshift the main effects of inhomogeneities on the luminosity distance of the sources of electromagnetic waves, such as standard candles, is the Doppler effect \cite{Bolejko:2012uj,Romano:2016utn}  due to the peculiar velocities of the sources and the observer, and a similar theoretical result holds also for the luminosity distance of GW sources \cite{Bertacca:2017vod}.
 It is consequently possible to apply the same reconstruction methods derived for standard candles \cite{Romano:2016utn,Chiang:2017yrq} to standard sirens, and perform a combined analysis of the luminosity distance of the sources of both  gravitational and electromagnetic waves.
At low redshift and in the perturbative regime the monopole of  the effects on the luminosity distance is proportional to the volume average of the density contrast \cite{Romano:2016utn}, and for an underdensity it  corresponds to an outwardly directed   peculiar velocity field pointing towards the outer  denser region, implying a local increase of the Hubble parameter, which could  account for the apparent difference between its  large and small scale estimation. 
 
Motivated by this apparent discrepancy we adopt an unbiased approach, i.e. we do not assume homogeneity, and let the data reveal whether the local Universe is in fact homogeneous or not. We analyze the luminosity distance data of supernovae (SNe) Ia from different catalogues combined with the luminosity distance of 9 GW sources identified by the Laser Interferometer Gravitational-Wave Observatory (LIGO) \cite{Abbott:2019yzh}, and use this combined set of data to reconstruct the peculiar velocity field of the sources. \hl{The redshift of GW events not having an electromagnetic transient counterpart was obtained \cite{Abbott:2019yzh} by the LIGO collaboration by applying  Bayesian analysis to the combined dataset of GW events and galaxy catalogues along the line of sight of the event.} 
%We first apply the inversion method using a fixed value of $\Omega_m$, to reconstruct the velocity field, and we then make other analyses fitting directly the luminosity distance including the effects of inhomogeneities and leaving $\Omega_m$ free to vary as well, to check its impact.
%We also analyze redshift shell averaged data, obtaining approximately the same results using no averaged data, showing that the impact of anisotropies is negligible, and that the effects of local structure  are dominated by the  monopole.

We perform a combined analysis of the  luminosity distance of SNe and GW sources, and we obtain different results using different SNe catalogues. 
We check if the peculiar velocity redshift correction applied to the Pantheon data could explain this difference, since Union 2.1 data is not redshift corrected, but we find that non redshift corrected Pantheon data lead to the same results.
We also perform some analyses using redshift shell averaged data, and obtain approximately the same results, hinting to the fact that the effects of the monopole component of the local inhomogeneity are the dominant ones.
%We find  a strong statistical evidence of a radial peculiar velocity field consistent with the one produced by a local underdensity with a density contrast of about $-0.48 \pm 0.003$, extending up to a redshift of about $0.083\pm 0.005$, in agreement with previous studies using number counts \cite{Keenan:2013mfa}.
The implications of the existence of a local underdensity  are profound, since not taking it into proper account can produce a mis-estimation of all background cosmological parameters obtained under the assumption of large scale homogeneity, and can explain for example the apparent discrepancy between different measurements of the Hubble constant \cite{Romano:2016utn}.

\section{Datasets}\label{sec: seleccion datos y coords}

\ch{We  analyze two different supernovae datasets, the Union 2.1 catalogue from the Supernova Cosmology Project (SCP) \cite{Union2}, and the Pantheon catalogue \cite{Scolnic_2018}, while for  GW sources we use data from the LIGO collaboration \cite{Abbott:2019yzh}.
The two different supernovae datasets lead surprisingly to very different conclusions in regards to the presence of a local underdensity. %Some inconsistencies in the Pantheon dataset were already found in \cite{Rameez:2019nrd,github},  precisely in relation to the redshift correction applied to the raw data, which is crucial  for the determination of the inhomogeneity characteristics. 

The formulae we use are including relativistic effects, as shown by comparison with exact numerical calculations in \cite{Romano:2016utn}, but if the peculiar velocity redshift correction is computed using an inaccurate formula often used to infer peculiar velocity from observations \cite{Davis:2014jwa}, the entire dataset can be affected. We used the second dataset published by the Pantheon collaboration \cite{Pantheon2}, which  removed  some   errors in the redshift correction at redshifts higher than the catalog depth. As explained in \cite{Romano:2016utn}, the 2M++ catalogue used to estimate the peculiar velocity redshift correction, is not deep enough to eliminate the effects of an inhomogeneity extending beyond its depth $z=0.067$. Note that the redshift edge of the inhomogeneity we obtain in our analysis is in fact  around the depth of 2M++. It should also be noted that the effects of the homogeneity extend slighly beyond the edge, as shown in one example in fig.(\ref{deltaz}).}

Using eq.(\ref{vDD}), taking $D_L(z)=D_L^{obs}(z)$ and $\overline{D}_L(z)$ as the luminosity distance of a $\Lambda CDM$ model with the parameters estimated by the Planck mission, we can obtain for each object the radial component of its peculiar velocity in a system of coordinates centered at our position. 
This procedure does not require to assume any spherical symmetry, since the general formula depends on the radial component of the peculiar velocity, and it only assumes that the dominant effect of the inhomogeneity is the Doppler effect, which is well justified at low redshift \cite{Romano:2016utn,Bolejko:2012uj},  $z<0.7$. Even using a cutoff of $z_{sup}=0.5$ we get the edge of the inhomogeneity to be $z_v < 0.09$, so the formula can be used safely in this range.

The radial velocity field is not necessarily spherically symmetric with respect to the observer position, since there can be anisotropies in the local structure, but here we will focus on the monopole component. There is evidence that local structure is not isotropic \cite{Romano:2014iea}, implying that extending the analysis to higher multipoles could also be important, but as a first step towards investigating the effects of local structure imprinted on the  luminosity distance we will focus on the monopole.
This can be done following two different approaches: computing  shell averages of redshift and velocity data before analyzing them, which are by construction isotropic, or  analyzing data without any averaging, assuming a model which includes the radial dependence, but ignores the possible angular dependence.
The averaging procedures consists in   dividing the datasets in spherical shells of constant width $\delta z=0.004$, and averaging the data inside each shell using inverse-variance weighting. 
We  analyze data following both methods, finding a small difference between the results, hinting to the fact that the effects of anisotropies are not strong.
 %This procedure has the advantage that the random component of the peculiar velocity field should be suppressed by this averaging procedure, but if a coherent radial component were present it should be manifest also in the monopole of the velocity. 
%We will denote as  $\bar{z}_i$ and $\bar{v}_i$ the error weighted shell averages of the redshift and peculiar velocities of the $i-th$ shell, and as $z_i$ and $v_i$ the redshift and peculiar velocity of each single supernova or GW source.
\ch{In the formulae defining the $\chi^2$ in the following sections, the variables $z_i$ and $v_i$ will represent the single data point or the shell average, depending on the fit.}

\section{Fitting the apparent magnitude data}
The observed quantity is the apparent magnitude $m$, and analyzing it directly is preferable for the Pantheon dataset, since $m$  is the quantity found in the publicly available data, not $D_L$, which is a derived quantity, and is model dependent, in the sense that it depends implicitly on $M$, which is a parameter of the model.
In fact, from the definition of distance modulus $\mu=m-M$, we get
\bea
D_L(z)&=&10^{\frac{\mu}{5}+1}=10^{\frac{m-M}{5}+1} \,,\label{DMU}
\eea
showing that in order to get $D_L^{obs}$ from $m^{obs}$ an assumption for $M$ has to be made. 
There are two main advantages in using $m$ :
\begin{itemize}
    \item It is not necessary to derive $D_L$ from $m$ and propagate the errors, as shown above.
    \item It is not necessary to use different datasets for different values of $\{H_0,M\}$, since these are just parameters of the model, while for $D_L$ they have to be assumed in order to obtain $D_L^{obs}$ from $m^{obs}$. This allows to plot and compare directly the results of the fits corresponding to different values of $\{H_0,M\}$.
\end{itemize}
Using $m$ it is clear the distinction between observed data and parameters of the model, while using $D_L$ the parameter $M$ is affecting both the model and $D_L^{obs}$, making the analysis less transparent. Theoretical predictions are normally made in terms of $D_L(z)$, but observational data analysis in models with  varying $\{H_0,M\}$ should be preferably performed in terms of $m$.
I<%n the case of Union 2.1 the published data is the distance modulus $\mu$, and only $H_0$ is given, so $M$ has to be obtained using the formula derived in the appendix B, to get $m$. 
%In this section we will nevertheless only focus on the Pantheon dataset.

The theoretical model for $m^{th}$ is related to the theoretical luminosity distance $D_L^{th}$ by
\be
m^{th}=5 \log{D_L^{th}}-5 + M \,,
\ee
and we compute the monopole effects of an hypothetical local inhomogeneity on $D_L$ using the fomula \cite{Romano:2016utn}
\be
D_L^{th}(z)=\overline{D_L^{th}}(z)\left[1+\frac{1}{3} f\overline{\delta^{th}}(z) \right] \,, \label{Dlowz}
\ee
where $\overline{\delta}(z)$ is the volume averaged density contrast and $\overline{D_L^{th}}(z)$ is the background luminosity distance of a $\Lambda CDM$ model. 

The dependency of the models on different parameters is given explicitly in the following equations
\bea
D_L(\Omega_i,H_0,I_i)&=&\overline{D_L}(\Omega_i,H_0)\left[1+\frac{1}{3} f\overline{\delta}(I_i) \right] \,,\\
m(\Omega_i,H_0,M,I_i)&=&5 \log{D_L(\Omega_i,H_0,I_i)}-5 + M \,,
\eea
where $I_i$ are the parameters modelling the inhomogeneity. A homogeneous model is a special case of the general model given above, with $I_i=0$. We do not make any assumption about the homogeneity of the local Universe, we just analyze data including the effects of a possible inhomogeneity. If the Universe were homogeneous our analysis should confirm it.

Note that as shown in \cite{Romano:2016utn,Chiang:2017yrq}, a  local inhomogeneity should only change the luminosity distance locally, since far from the inhomogeneity the volume averaged density contrast of a finite size homogeneity tends to zero.

\section{Modeling the local inhomogeneity}
\begin{figure}[H]
\centering
 \includegraphics[scale=0.37]{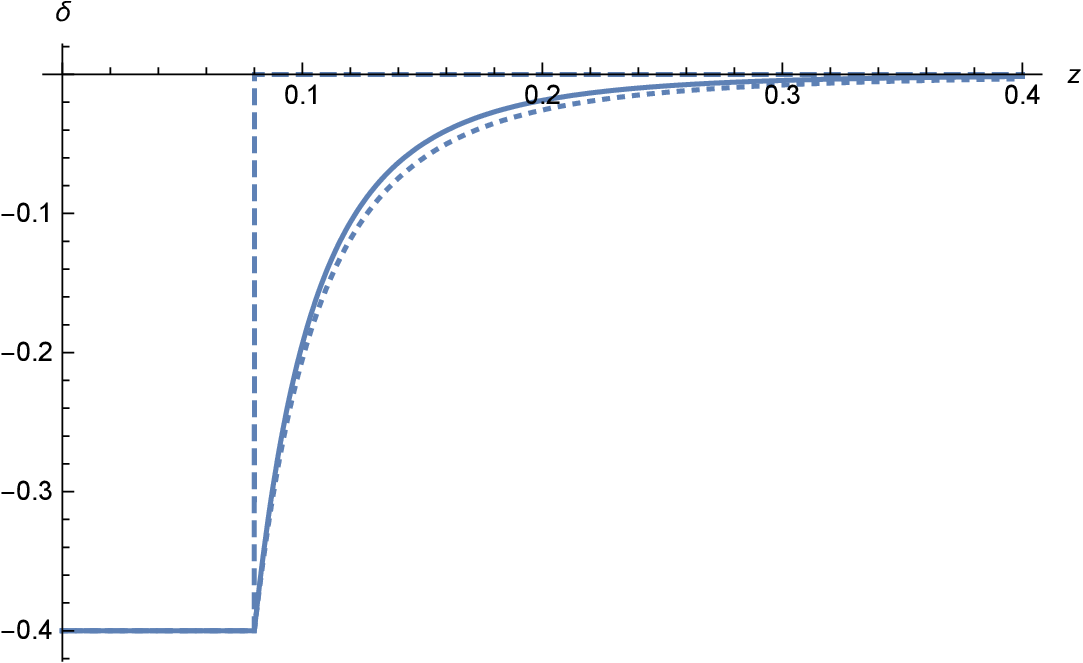}
 \caption[]{\ch{The step density contrast $\delta$ (dashed) defined in eq.(\ref{deltastep}), its volume average $\overline{\delta}$ (solid) obtained in eq.(\ref{eq: delta ave piecewise}), and the approximation (dotted) used to derive  eq.(\ref{eq: trazoz_lineal_cub}) are plotted as a function of redshift for $z_v=0.08$ and $\delta_v=-0.4$. The effects of the inhomogeneity are proportional to $\overline{\delta}$, and  extend beyond the edge of the void, but are quickly suppressed beyond the edge, implying that high redshift observations are not affected by the local inhomogeneity. }}
 \label{deltaz}
\end{figure}

We model the local underdensity with a density profile of the type
\be
\delta^{th}(\chi)=\delta_v [1-\theta(\chi-\chi_{v})] \,, \label{deltastep}
\ee
where $\chi_v$ is the comoving distance of the edge of the inhomogeneity, $\delta_v$ the density contrast inside the inhomogeneity, and $\theta(x)$ is the Heaviside function.
The corresponding volume averaged density contrast is
\begin{equation}
\overline{\delta^{th}}(z)=
     \left\{
     \begin{array}{lcc}
     \delta_{v} &  & z<z_{v}  \, \\ 
     \delta_{v} \left[ \frac{z_v(1+z_v)}{z(1+z)} \right]^3   && z>z_{v} \,
     \end{array}
     \right \} \,,
\label{eq: delta ave piecewise}
\end{equation}
where $\delta_v$ is the density contrast inside the inhomogeneity and $z_{v}$ is the inhomogeneity edge redshift. The derivation of this formula is given in the appendix C.
Using eq.(\ref{vrd}) we get the following formula for  the radial velocity profile

\begin{equation}
\frac{v_r(z)}{c}=\frac{1}{3}f 
\left\{
     \begin{array}{lcc}
      \delta_v\,z & for & z<z_{v} \\
      d \, /z^{2}  &for& z>z_{v} 
     \end{array}
\right. \,,
\label{eq: trazoz_lineal_cub}
\end{equation}
where the matching condition at $z_v$ implies $d=\delta_v \, z_v^3$.
\ch{Note that in deriving the last equation we have neglected the factor $(1+z_v)/(1+z)$, which is a good approximation, as shown in fig.(\ref{deltaz}). The above formula is in agreement with the general result obtained in \cite{Romano:2016utn}, confirmed by exact numerical calculations, according to which for low redshift inhomogeneities the effects are suppressed at high redshift by the volume in the denominator of the volume average.

We fit the data with this model by minimizing with respect to the two parameters $\delta_v,z_v$ the following $\chi^2(\delta_v,z_v)$
%\medskip
\begin{equation}
\chi^2 = \sum_i
       \left[
       \frac{m_i- m^{th}(z_i)}{\sigma_{m_i}}
       \right]^2 \,.
\label{eq: chi2: m}
\end{equation}
The results of the fit using the covariance matrix are given in the appendix.
%\medskip

The difference between our analysis and the one reported in \cite{Kenworthy:2019qwq} is presumably due to the fact therein it was assumed a fixed value of the absolute magnitude M, ignoring how it should change if $H_0$ is different, and instead of fitting directly the apparent magnitude, the intercept $a_b$ was used, whose mathematical definition relies implicitly on some homogeneity assumptions. 
The difference between our analysis and \cite{Kenworthy:2019qwq} cannot be attributed to a difference in the choice of radial coordinate, since the luminosity distance as function of the redshift is an observational quantity, and such it is invariant under any coordinate transformation. }

The above model is based on the very accurate  low redshift approximation derived in \cite{Romano:2016utn}, and on the assumption of a constant step density contrast profile as given in eq.(\ref{deltastep}). Note that since $\overline \delta(z)$ is the volume average of the density contrast, outside the inhomogeneity it is inversely proportional to the volume, and since at low redshift $\chi \approx z/a H_0 $, we get the above expression outside the inhomogeneity, which is proportional to the inverse cube of the redshift. The factor $(1+z)$ comes from the scale factor in $\chi \approx z/a H_0$, and at low redshift can be safely neglected. As shown in \cite{Romano:2016utn}, by explicit comparison with exact numerical results, these approximations are quite accurate at low redshift.

It is important to note that the gravitational effects of the inhomogeneity extend slightly beyond its edge, due to the inverse cube suppression. 

%using the $M,H$ published in \cite{Riess:2016jrr} as reference.
\section{Fitting consistently data with models assuming different value of $H_0$ and $M$}
Using only the definition of distance modulus and eq.(\ref{DMU}) we derive in Appendix B a general model independent relation between different values of $\{H_0,M\}$ estimated at low redshift
\bea
M_{a} &=& M_{b} + 5\log_{10}\left(\frac{H_{a}}{H_{b}}\right) \,,
\eea
which allows, for example, to obtain the implied Planck value of $M^p$ from $H_0^P$ and $\{H_0^R,M^R\}$, which are the values estimated in \cite{Riess:2016jrr}. In the Appendix B we discuss the limits of validity of this relation, but when considering models with the same parameters $\Omega_i$ it can be safely applied, or more in general but at low red-shift, where $\{H_0,M\}$ are estimated. Note that a similar relation for $M$ was used in \cite{Chiang:2017yrq} and more recently in \cite{Benevento:2020fev}. There it was claimed it to be valid only for $\Lambda CDM$ models, while our derivation is completely model independent.
Using this relation, and assuming the values obtained in \cite{Riess:2016jrr} as reference, we obtain $M^P$ and fit $m^{obs}$ with  different homogeneous and inhomogeneous models.%, with different values of $\{H_0,M\}$.

We denote with $m^{\textup{Hom}}(H_0^R)$ and $m^{\textup{Inh}}(H_0^P)$ respectively a homogeneous model with $\{H_0,M\}=\{H_0^R,M^R\}$ and an inhomogeneous model with $\{H_0,M\}=\{H_0^P,M^P\}$.
We consider flat cosmologies with a $\Omega_m$ value estimated from the  Planck mission data \cite{Abbott:2019yzh}.

We fit data across different redshift ranges, to verify the robustness of the results.
The results are summarized in Table \ref{tab: fit m + no cov + inhP + homR + 0.11,0.5,1.5 }, and shown in fig.(\ref{fig: Pantheon mVsz + void + homo + Planck + Riess, no shells, zsup=0.11}-\ref{fig: Pantheon mVsz + void + homo + Planck + Riess, no shells, zsup=0.5}). As can be seen the $\chi^2$ of $m^{\textup{Hom}}(H_0^R)$ is higher than that of $m^{\textup{Inh}}(H_0^P)$  for $z_{sup}=0.11$, i.e. a model with $H_0^P$ can fit better the low redshift SNe when the effects of a local small inhomogeneity are included in the analysis. For higher cutoffs the two models have approximately the same $\chi^2$. 

Note that the density contrast of the best fit underdensity is not large, and the hedge is located around the depth of the catalog $2M++$ used in \cite{Riess:2016jrr} for the redshift correction, supporting the argument \cite{Romano:2016utn} that the apparent tension between $H_0^P$ and $H_0^R$ is the consequence of a local inhomogeneity whose effects have not been removed by redshift correction, due to its size being comparable to  the $2M++$ depth.
As explained in \cite{Romano:2016utn} and evident from the plots, the local inhomogeneity does not affect the luminosity distance at high redshift, since the effect is proportional to the volume average of the density contrast, which tend to zero at high redshift, for a finite size inhomogeneity.

This kind of inhomogeneity could have been  generated by a  peak of primordial curvature perturbations \cite{Romano:2014iea}, which is not a very unlikely event. Independently from the probability of being located inside of such an inhomogeneity, the unbiased data analysis we have performed is giving  statistical evidence of its presence. This should be considered for its statistical significance more than for the theoretical estimation of its probability, i.e. the existence of inhomogeneities should be tested using observational data rather then being excluded a priori from the analysis, on the basis of theoretical predictions. %, since the latter can rely on numerical simulations involving several parameters or assumptions whose implications are difficult to track or verify, while our analysis is based on very transparent assumptions and can be easily reproduced.
Most of the GW sources are located outside this inhomogeneity, explaining why the $H_0$ value estimated from these  objects is in agreement with the estimation based on other large scale observations such as the CMB.
 %%?? compare Gw and not GW to see if it is cause of the difference between m and v fit. specify where GW is not included. estimate prob using matter power spectrum fit

\begin{figure}[H]
    \centering
    \includegraphics[scale=0.35]{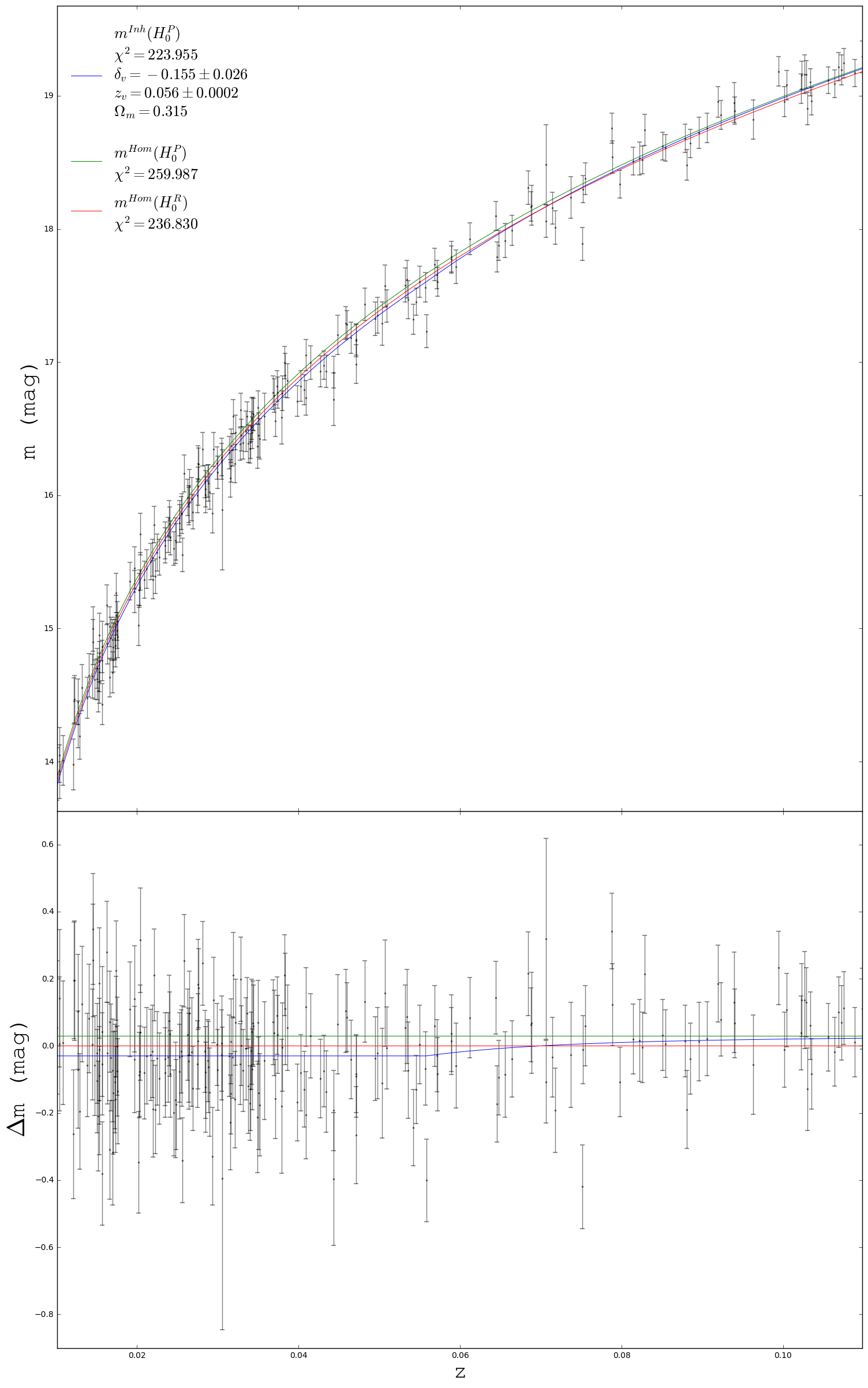}
    \caption{Fit of Pantheon dataset with different models, for SNe with $z<0.11$. In the lower panel is plotted for comparison purposes the difference between $m$ and $m^{\textup{Hom}}(H_0^R)$, $\Delta m=m-m^{\textup{Hom}}(H_0^R)$. While $m^{\textup{Hom}}(H_0^P)$ does not provide a good fit of the data, $m^{\textup{Inh}}(H_0^P)$ is fitting the data better than $m^{\textup{Hom}}(H_0^R)$.}
    \label{fig: Pantheon mVsz + void + homo + Planck + Riess, no shells, zsup=0.11}
\end{figure}

\begin{figure}[H]
    \centering
    \includegraphics[scale=0.35]{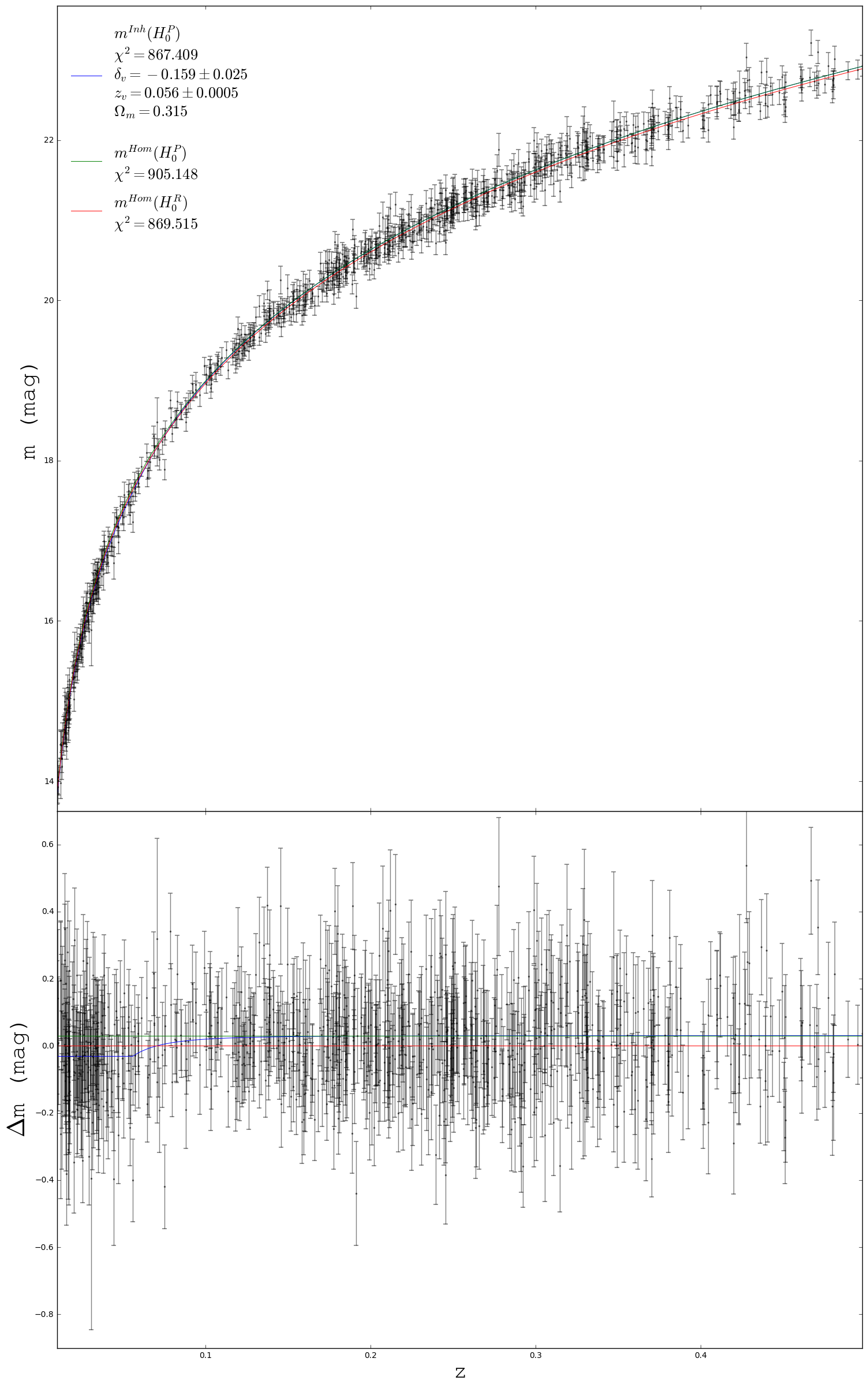}
    \caption{Fit of Pantheon dataset with different models, for SNe with $z<0.5$. In the lower panel is plotted for comparison purposes the difference between $m$ and $m^{\textup{Hom}}(H_0^R)$, $\Delta m=m-m^{\textup{Hom}}(H_0^R)$. While $m^{\textup{Hom}}(H_0^P)$ does not provide a good fit of the data, the $\chi^2$ of $m^{\textup{Inh}}(H_0^P)$ is approximately the same of $m^{\textup{Hom}}(H_0^R)$.}
    \label{fig: Pantheon mVsz + void + homo + Planck + Riess, no shells, zsup=0.5}
\end{figure}

\begin{table}[H]
\centering
\resizebox{17cm}{!} {
\begin{tabular}{lccc c ccc c c}
\hline
 & \multicolumn{3}{c}{$z_{sup}=0.11$} && \multicolumn{3}{c}{$z_{sup}=0.5$} && \multicolumn{1}{c}{$z_{sup}=1.5$} \\
\cline{2-4} \cline{6-8} \cline{10-10}
 & $\delta_v$   & $z_{v}$  & $\chi^2$  & &  $\delta_v$ & $z_{v}$ & $\chi^2$  & &  $\chi^2$    \\
\cline{2-4} \cline{6-8} \cline{10-10}

$m^{\textup{Inh}}(H_0^P)$ & $-0.155 \pm 0.026 $ & $0.056 \pm 1.84\times 10^{-4} $ &  223.955 & & $-0.159 \pm 0.025 $ & $0.056 \pm 4.6\times 10^{-4} $ & 867.409  & &  1036.026  \\
$m^{\textup{Hom}}(H_0^P)$ & -                   & -                               &  259.987 & & -                   & -                              & 905.148  & &  1073.773  \\
$m^{\textup{Hom}}(H_0^R)$ & -                   & -                               &  236.830 & & -                   & -                              & 869.515  & &  1040.865  \\

\hline
\end{tabular}
                  } 
    \caption{Results of the fits of the Pantheon data with different upper limits $z_{sup}$ of the redshift range. An inhomogeneous model $m^{\textup{Inh}}(H_0^P)$ can fit the data better than a homogeneous model $m^{\textup{Hom}}(H_0^R)$ for $z_{sup}=0.11$, while for higher cutoffs the $\chi^2$ is approximately the same. This resolves the apparent $H_0$ tension, since at high redshift the local inhomogeneity does not affect the luminosity distance. The $\chi^2$ for $z_{sup}=1.5$ is obtained using the best fit parameters obtained for the $z_{sup}=0.5$ fit.}
    \label{tab: fit m + no cov + inhP + homR + 0.11,0.5,1.5 }
\end{table}

\section{Another approach to the reconstruction of the  peculiar velocity field from the luminosity distance}
Assuming the effects of the observer velocity have been removed, the dominant effect of inhomogeneities on the angular diameter distance at low redshift \cite{Hui:2005nm} is given by the radial component of the emitter peculiar velocity $v_e$

\begin{equation}
D_L(z) \approx \overline{D}_L(z)\left[1+\frac{v_e\cdot \textbf{n}}{z\, c}\right] \,, \label{Ddz}    
\end{equation}
where $c$  is the speed of light, the unit vector $\textbf{n}$ is in the direction of propagation from the emitter to the observer, and $\overline{D}_L(z)$ is the background angular diameter distance. In the above equation we have assumed that the peculiar velocity of the observer is zero, which is consistent with analyzing data from which our peculiar velocity with respect to the cosmic microwave background (CMB) has been subtracted, as is the case for the SNe Ia datasets we consider.

The background angular diameter distance $\overline{D}_L(z)$ is predicted theoretically for each value of $z$  using the background cosmological parameters measured by large scale observations of the Universe, in our case the CMB measurements of the Planck \cite{Akrami:2018odb,Aghanim:2018eyx} mission, while $D_L(z)$ is the observed distance. It is then immediate to determine the radial component $v_r$ of the peculiar velocity  from eq.(\ref{Ddz})

\be
 v_r= -v_e \cdot \textbf{n}= -z\, c \left(\frac{D_{L }}{\overline{D}_L} - 1 \right)=-z\, c \delta D_L \,,
\label{vDD}
\ee
where we introduced the luminosity distance contrast $\delta D_L$, a  dimensionless quantity which accounts for the relative difference between the observed luminosity distance $D_L$ and the corresponding background value $\bar{D}_L$. Note that the relation  $ v_r= -v_e \cdot \textbf{n}$ is due the opposite direction of $\textbf{n}$ with respect to the outwardly directed  radial coordinate centered at the observer. Eq.(\ref{vDD}) is the foundation of the reconstruction method we will apply, and allows to find the radial peculiar velocity field of the sources from the difference between their measured luminosity distance and  the corresponding theoretical prediction obtained assuming a Friedman model with cosmological parameters obtained from independent large scale observations which are not sensitive to local inhomogeneities, such as the CMB \cite{Akrami:2018odb,Romano:2016utn}.

The monopole of the effects of local inhomogeneities can be computed by choosing a spherical coordinate system centered at the observer position, and after integrating the Euler's equations we get \cite{peebles:1993} 

\begin{equation}
\frac{v_{r}(z)}{c} = -\frac{1}{3}f\bar{\delta}(z)z  \,,
\label{vrd} 
\end{equation}
where $f=\frac{1}{H}\frac{\dot{D}}{D}$ is the growth factor and the volume averaged density contrast is defined as

\bea
\overline{\delta}(\chi)&=&\frac{3}{4 \pi \chi^3}\int^{\chi}_0 4 \pi \chi'^2 \delta(\chi') \, \label{dav} d\chi' \,.
\eea
\hl{
 It has been shown in  \cite{Romano:2016utn} (see fig.(2) and section 6 therein)  that at low redshift the linear approximation adopted to obtain the above equations is in good agreement with exact numerical calculations in the redshift range of interest in this paper.}
Eq.(\ref{vrd}) is the basis of our analysis since it allows to model appropriately the peculiar  velocity  field in terms of the density contrast. An underdensity induces an outwardly oriented velocity field, which if not distinguished from the large scale expansion due to the Hubble flow, can lead to an apparent discrepancy between the measurement of the Hubble constant  obtained from local and large scale observations  \cite{Romano:2016utn}.
An inversion method to determine the  density contrast from the luminosity distance contrast was derived in \cite{Romano:2016utn}, but it involves derivatives  with respect to the redshift, making it difficult to apply to observational data, while the peculiar velocity obtained using eq.(\ref{vrd}) is more suitable for data analysis since it does not involve any derivative, and for this reason it is more convenient for data analysis purposes.

\section{Fitting the peculiar velocity}
\begin{figure}[H]
\centering
 \includegraphics[scale=0.33]{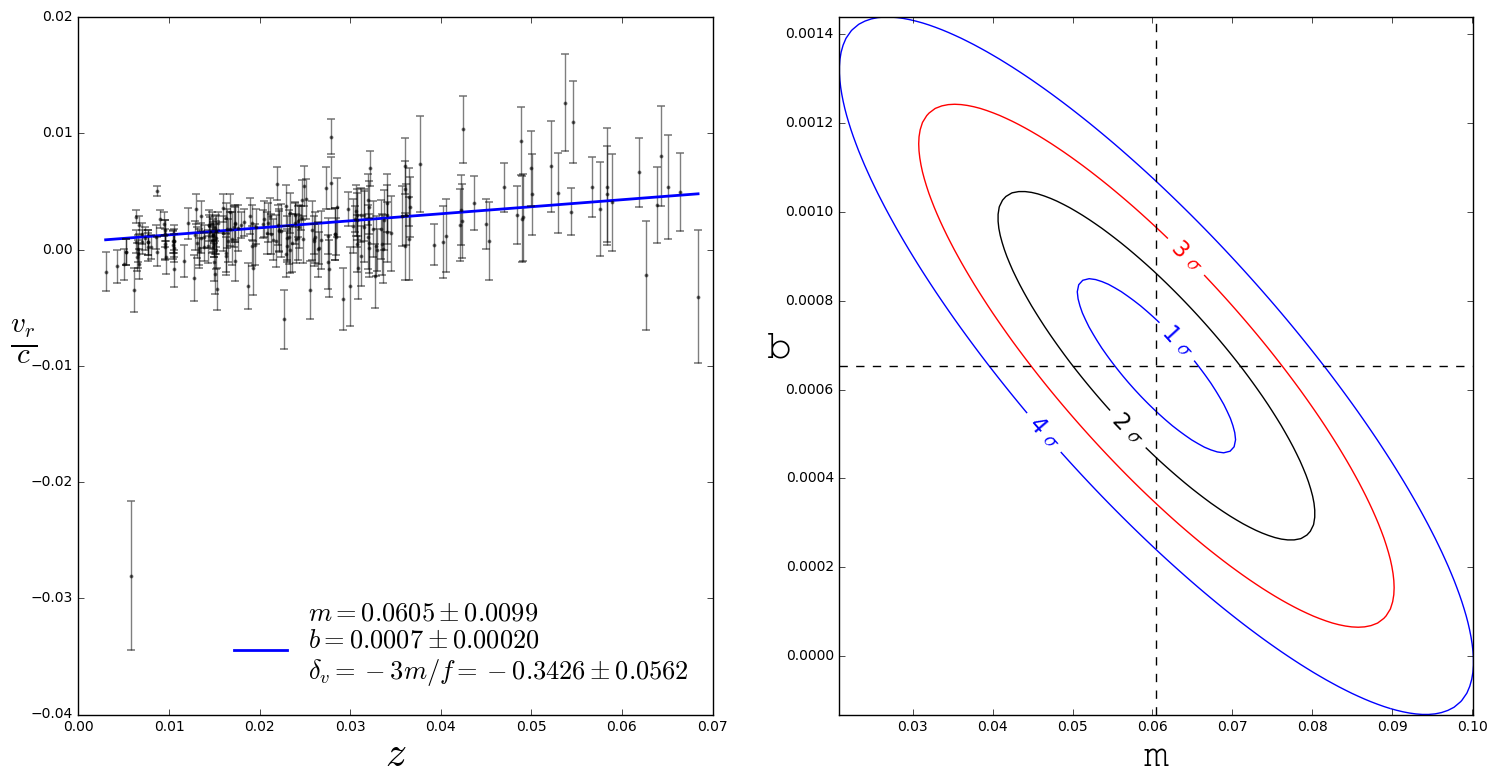}
    \caption[]{Results of the fit of eq.(\ref{eq: chi2: vr y z para SN}) with not shell averaged data for the Union 2.1 data, and maximum redshift $z_{sup}=0.0685$. We can exclude the $m=0$ null hypothesis at more than $4\sigma$ confidence level. }
 \label{fig:nohedge-noshell}
\end{figure}

\begin{figure}[H]
\centering
 \includegraphics[scale=0.33]{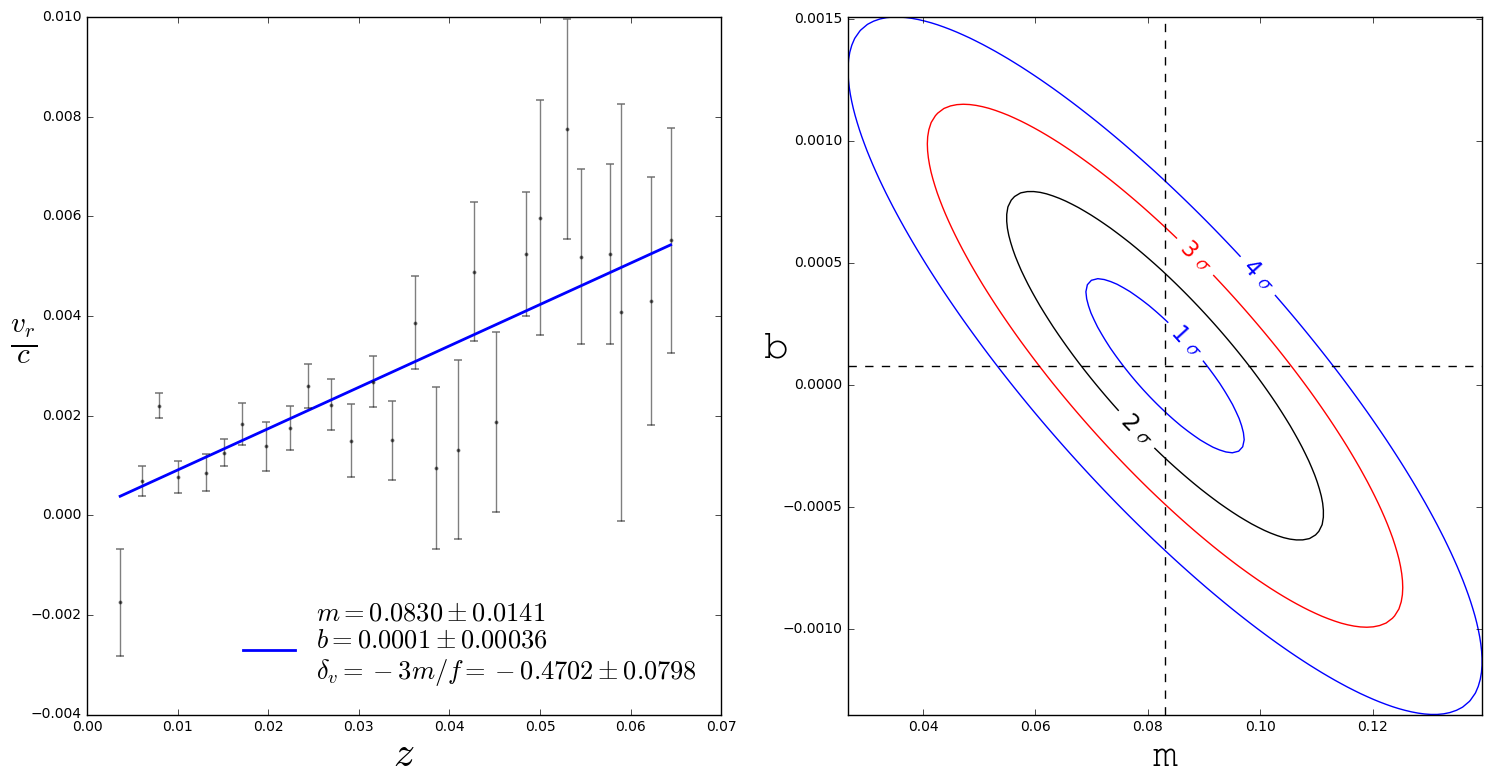}
 \caption[]{Results of the fit of eq.(\ref{eq: chi2: vr y z para SN}) with shell averaged data for the Union 2.1 data, and maximum redshift $z_{sup}=0.0685$. We can exclude the $m=0$ null hypothesis at more than $4\sigma$ confidence level. The shells have width $\Delta z=0.004$.}
 \label{fig:nohedge-shell}
\end{figure}

In applying the reconstruction method we will assume the background cosmological parameters obtained by the Planck mission \cite{Abbott:2019yzh}.
According to eq.(\ref{vrd}) the radial velocity due to the monopole of the density contrast should be given by $v_{r}(z)/c= - \frac{1}{3}f\bar{\delta}(z)z$, which inside a region with constant density contrast $\delta_v$ gives $\bar{\delta}(z)= \delta_v$ and $v(z) \propto cz $. %This shows that the Hubble parameter can be affected by a local inhomogeneity, and in particular that an underdensity can induce a local increase of the Hubble parameter \cite{}.
Motivated by the above considerations, we will fit the monopole of the low redshift peculiar velocity field inferred from luminosity distance observation assuming a linear relation of the form  $v = m\, z + b$, and obtain the best fit parameters by minimizing the $\chi^2$  defined as 

\begin{equation}
\chi^2=\sum_{i=1}^{N}
       \left[
       \frac{v_i-(m \,z_i+b)}
            {\sqrt{\sigma_{v_i}^2 + \sigma_{z_i}^2}}
       \right]^2 \,.
\label{eq: chi2: vr y z para SN}
\end{equation}
%where $\bar{v}_i$ and $\bar{z}_i$ are the shell averages of velocity and redshift, and $\bar{\sigma}_{v_i}$ and $\bar{\sigma}_{z_i}$ their errors, obtained by Gaussian propagation of the errors of the values of the elements of each shell.
The parameter  $m$ is related to the volume averaged density contrast  $\bar{\delta}$ by the relation $\bar{\delta}(z)=-\frac{3 m}{f c}$.
\ch{The results for the Union data fits are shown in fig.(\ref{fig:nohedge-shell}) and fig.(\ref{fig:nohedge-noshell}). For both the averaged and the not averaged data,  there is a strong statistical evidence of the presence of a radial velocity field directed outwardly, which corresponds to a volume averaged density contrast of about $\delta_v=-0.4$. The difference between the two fits, one with  shell averaged data and the other with single data points, is a hint to the importance of anisotropies in this redshift range, i.e. for $z_{sup}=0.0685$, but since this difference is not so large, it implies the monopole is producing the dominant effect.}
%In our analysis we will use the Union 2.1 compilation dataset \cite{N.Suzuki:2012}
%% 5, vsn=m z+b, SN
%% 11, vsn= alpha z2m+ v2m +gamma
%% 12  vsn= mcor v2m+bcor
%% 13,15 vsn= alpha z, beta/z^3

\section{Determining the size of the inhomogeneity}
We have shown that low redshift observations support the existence of a local underdensity, but it is also important to determine its size. 

\begin{center}
\begin{figure}[H]
 \centering
 \includegraphics[scale=0.33]{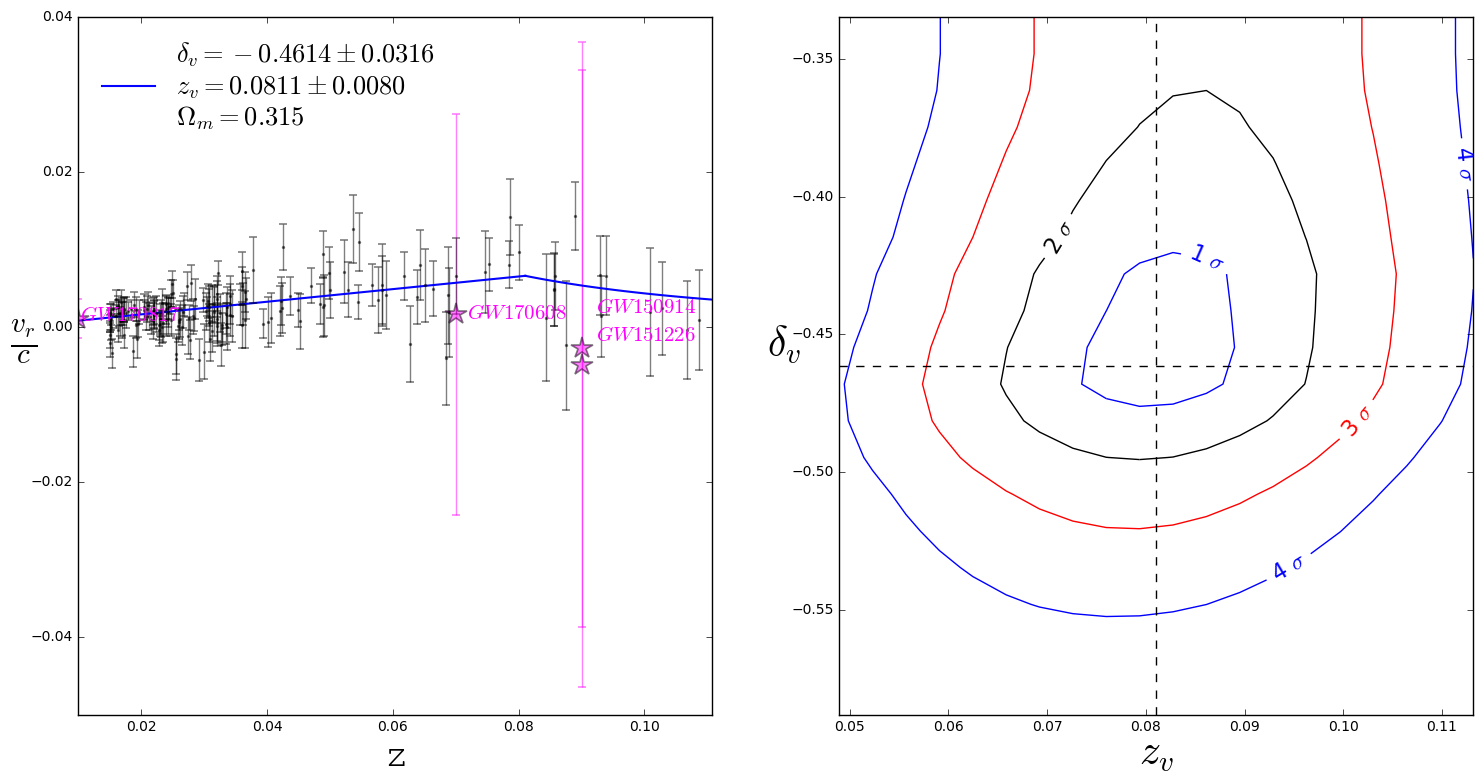}
 \caption[]{Results of the fit of eq.(\ref{eq: chi2: funcion a trazos}) with not shell averaged data for the Union 2.1 data, and maximum redshift $z_{sup}=0.11$. We can exclude the $\delta_v=z_v=0$ null hypothesis at more than $4\sigma$ confidence level. The fitted data is a combination of the GW sources and SNe.}
 \label{fig:hedge-noshell}
\end{figure}
\end{center}

\begin{figure}[H]
 \centering
 \includegraphics[scale=0.33]{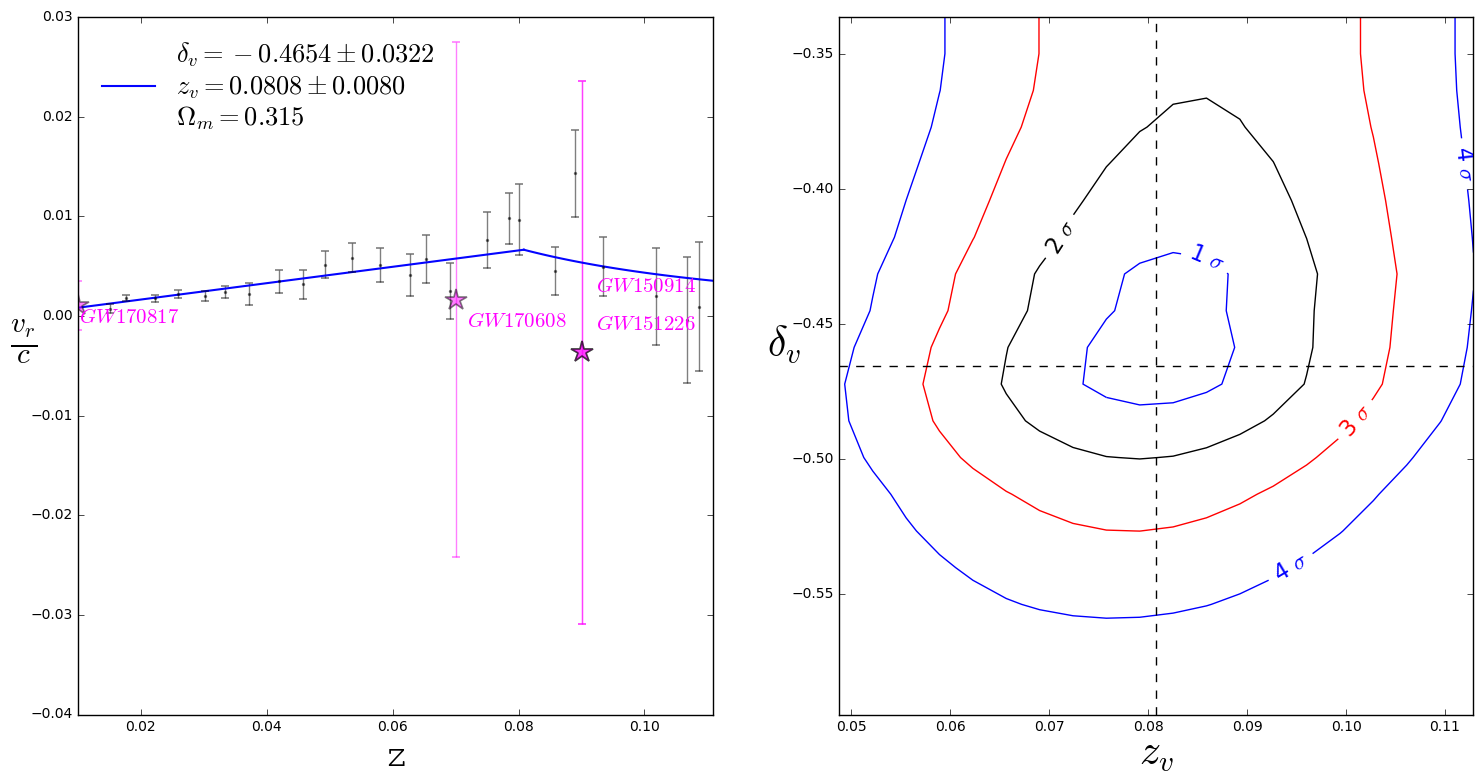} 
\caption[]{Results of the fit of eq.(\ref{eq: chi2: funcion a trazos}) with shell averaged data for the Union 2.1 data, and maximum redshift $z_{sup}=0.11$. We can exclude the $\delta_v=z_v=0$ null hypothesis at more than $4\sigma$ confidence level. The shells have width $\Delta z=0.004$, and the averages have been obtained by combining SN and GW sources data in the same shell. The fitted data is a combination of the GW sources and SNe.} 
\label{fig:hedge-shell}
\end{figure}

\begin{center}
\begin{figure}[H]
 \centering
 \includegraphics[scale=0.33]{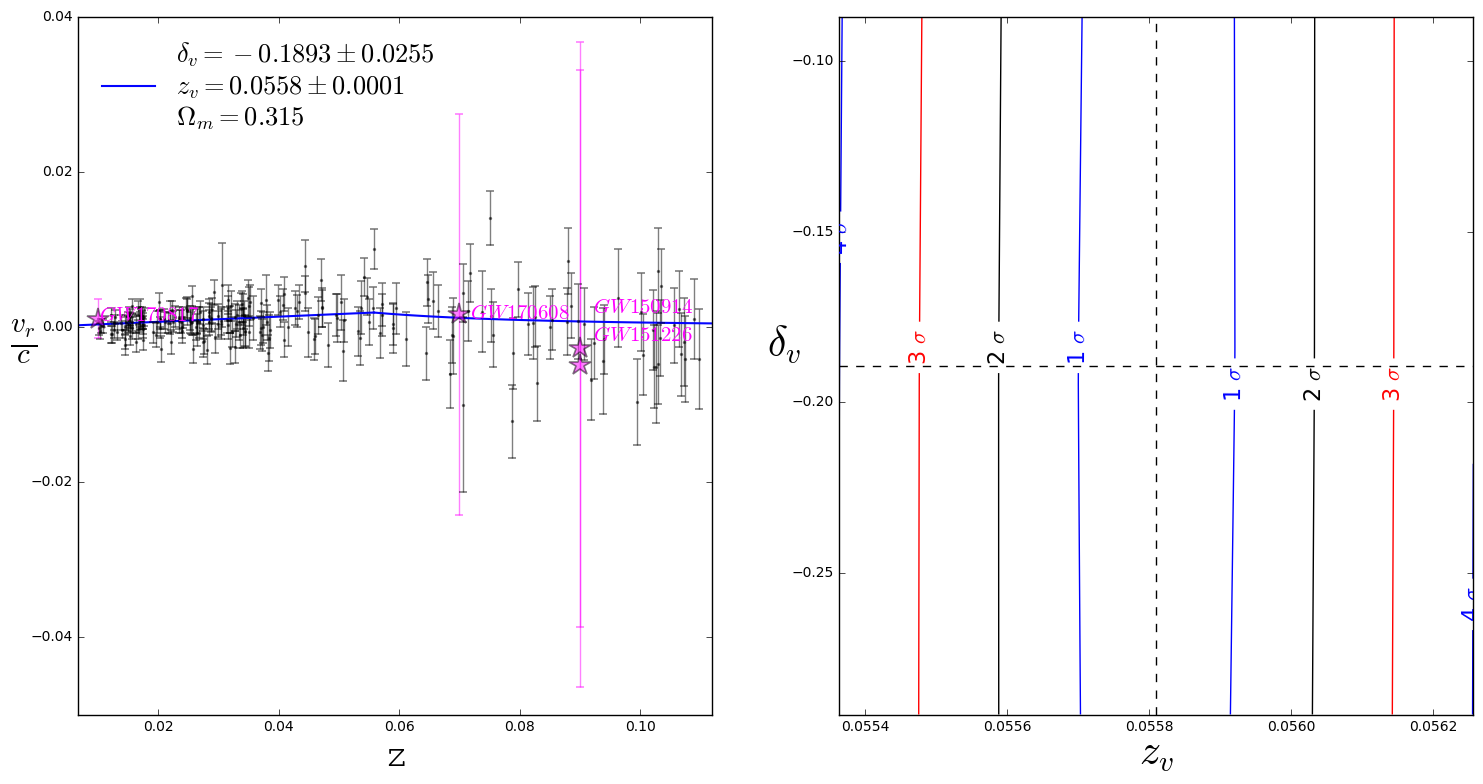}
 \caption[]{Results of the fit of eq.(\ref{eq: chi2: funcion a trazos}) with not shell averaged data for the Pantheon data, and maximum redshift $z_{sup}=0.11$.  We can exclude the $\delta_v=z_v=0$ null hypothesis at more than $4\sigma$ confidence level. The fitted data is a combination of the GW sources and SNe.}
 \label{fig:Pantheon hedge-noshell}
\end{figure}
\end{center}

%\section{Analysis with the luminosity distance $D_L$}

\begin{figure}[H]
    \centering
    \includegraphics[scale=0.6]{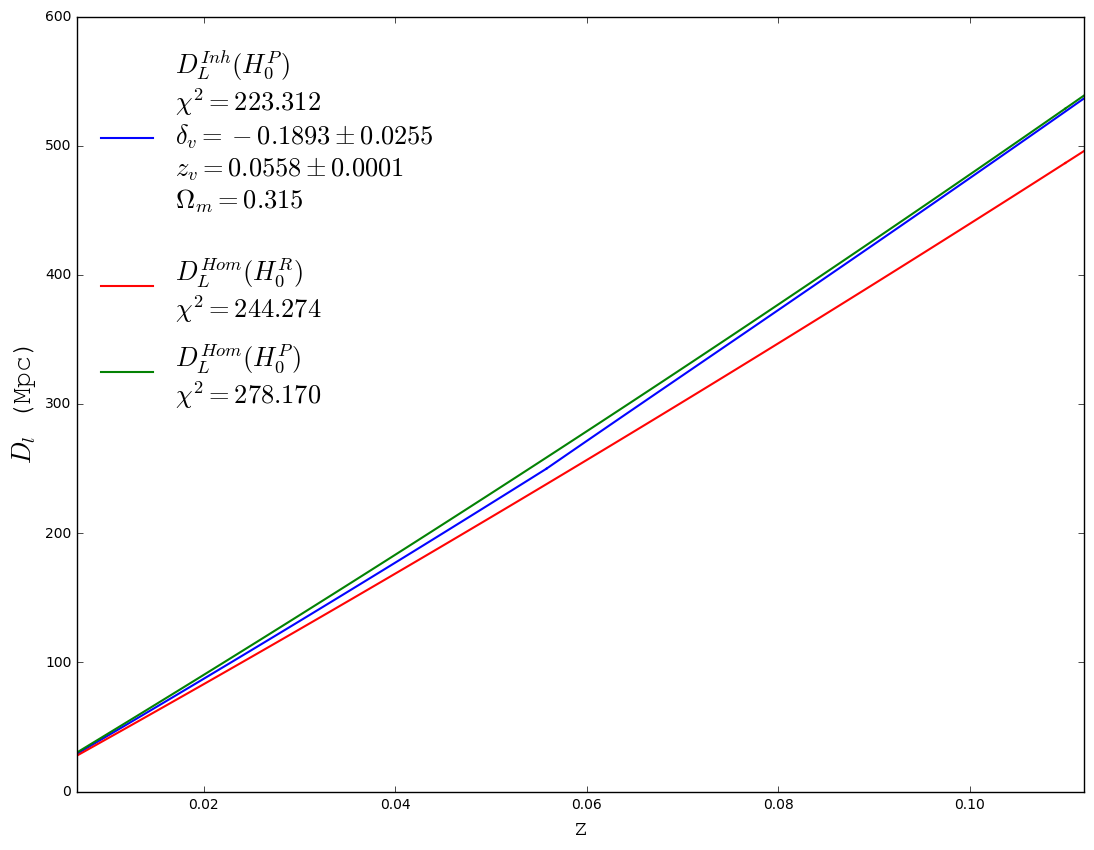}
    \caption{Best fit inhomogeneous model $D_L^{\textup{Inh}}(H^P_0)$ obtained using the reconstruction method with the Pantheon data up to a redshift $z_{sup}=0.11$. We also show the luminosity distance of the homogeneous models $D_L^{\textup{Hom}}(H^P_0)$ and $D_L^{\textup{Hom}}(H^R_0)$ for comparison purposes. The luminosity distance $D_L^{\textup{Inh}}(H^P_0)$ at high redshift tends as expected to $D_L^{\textup{Hom}}(H^P_0)$, implying that the inhomogeneity does not affect the large scale estimation of $H_0$, while at low redshift it is close to  $D_L^{\textup{Hom}}(H^R_0)$.}
    \label{fig: Pantheon Dl + voidP + homo PR + no shells + z_sup=0.11}
\end{figure}

\begin{figure}[H]
 \centering
 \includegraphics[scale=0.33]{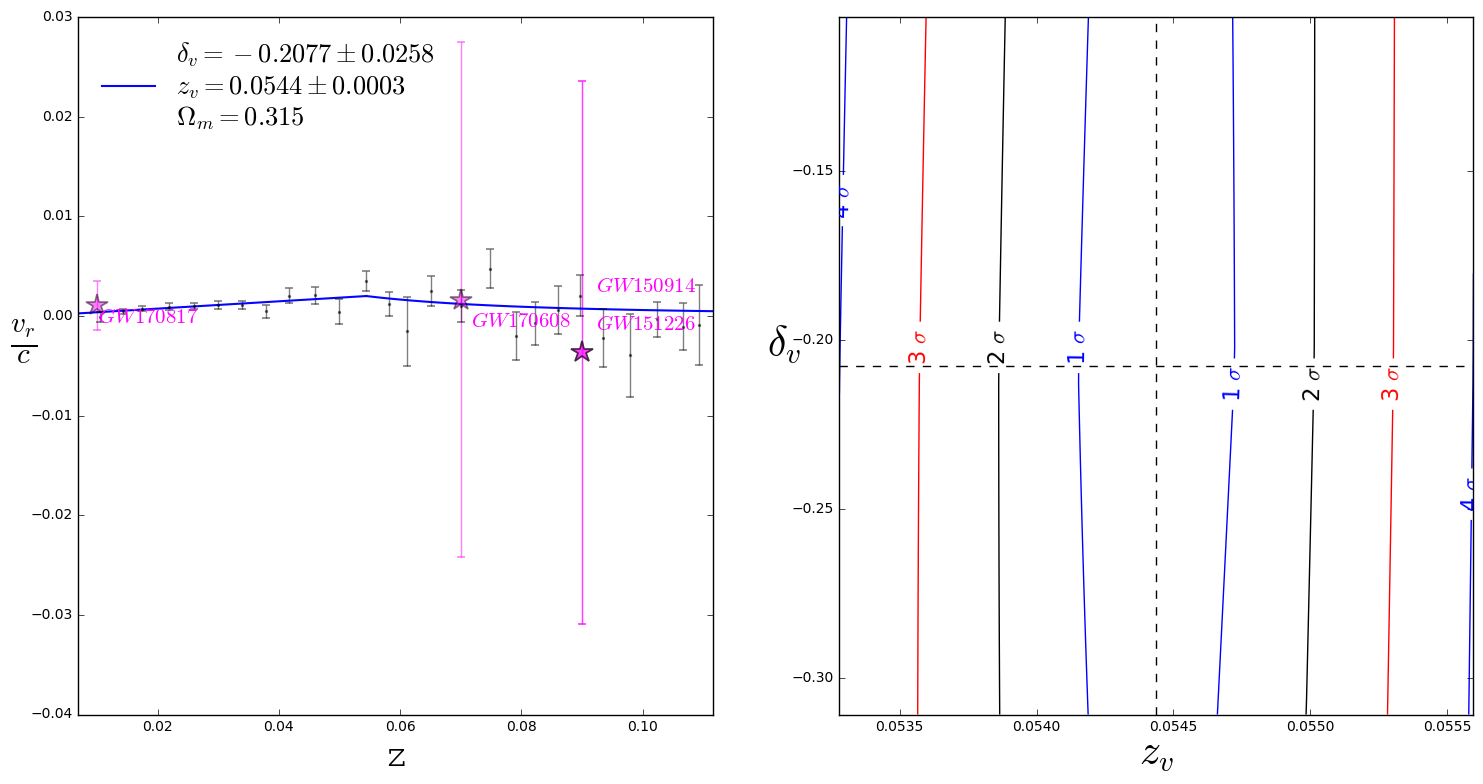} 
\caption[]{Results of the fit of eq.(\ref{eq: chi2: funcion a trazos}) with shell averaged data for the Pantheon data, and maximum redshift $z_{sup}=0.11$. The shells have width $\Delta z=0.004$, and the averages have been obtained by combining SNe and GW sources data in the same shell.  We can exclude the $\delta_v=z_v=0$ null hypothesis at more than $4\sigma$ confidence level. The fitted data is a combination of the GW sources and SNe.} 
\label{fig:Pantheon hedge-shell}
\end{figure}

We fit the data with this model by minimizing with respect to the two parameters $\delta_v,z_v$ the following $\chi^2(\delta_v,z_v)$

\medskip
\begin{equation}
\chi^2 = \sum_{z<z_v}
       \left(\frac{v_i- \frac{1}{3}f\delta_v z_i}
            {\sigma_{v_i} } 
       \right)^2 
       +
       \sum_{z>z_v}
       \left(\frac{v_i- \frac{1}{3z_i^2}f\delta_v z_v^3}
            {\sigma_{v_i} } 
       \right)^2 \,.
\label{eq: chi2: funcion a trazos}
\end{equation}
\medskip

We perform two different fits, one without shell averaged data,  and another with shell average, and  the results are shown respectively in fig.(\ref{fig:hedge-noshell}) and fig.(\ref{fig:hedge-shell}) for the Union 2.1 data and in fig.(\ref{fig:Pantheon hedge-noshell}-\ref{fig: Pantheon Dl + voidP + homo PR + no shells + z_sup=0.11}) and fig.(\ref{fig:Pantheon hedge-shell}) for the Pantheon data. As explained previously, depending on the type of fit the subindex ${}i$  corresponds respectively to shell averages or to the single data points.
As can be seen in the figures, the results of the fit using shell averaged data or single data points are very similar, hinting to the fact that the effects of anisotropies are not very strong.

The confidence contour plots show for both datasets the presence of a radial velocity field corresponding to a local underdensity, but the best fit parameters are quite different. This could be due to the different sky coverage of the two datasets, or the difference in the light curve fitting methods.
The results of the Pantheon dataset are similar to the ones obtained fitting the apparent magnitude, showing that the inversion method is working well.
Nevertheless the direct fit of the apparent magnitude is preferable since it does not require to perform any error propagation for $D_L$, which can affect the accuracy of the inversion results.

\section{Impact of the gravitational waves data on the fit results}
\begin{center}
\begin{figure}[H]
 \centering
 \includegraphics[scale=0.37]{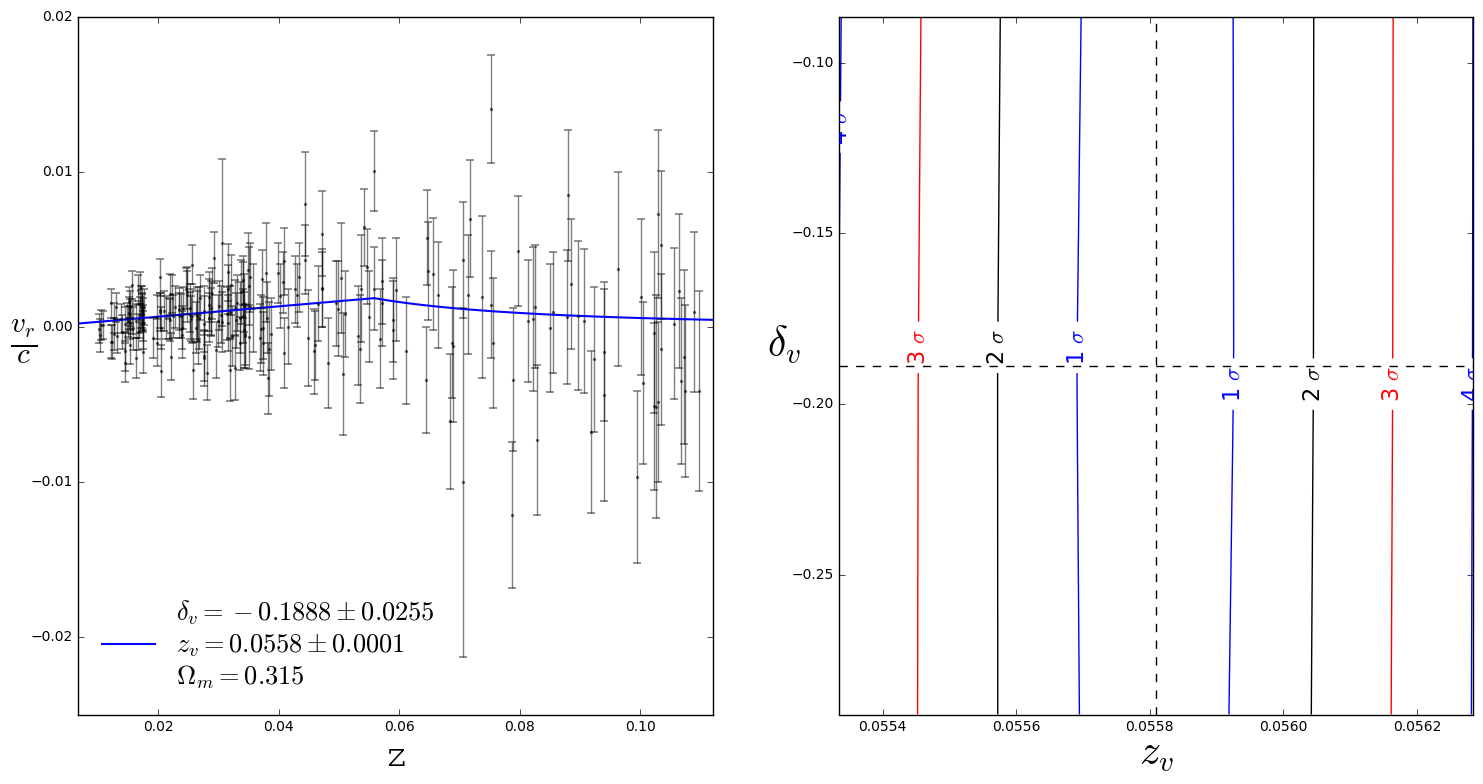}
 \caption[]{Results of the fit of eq.(\ref{eq: chi2: funcion a trazos}) with not shell averaged data for the Pantheon data, for a maximum redshift of $z_{sup}=0.11$, and without including GW data. We can exclude the $\delta_v=z_v=0$ null hypothesis at more than $4\sigma$ confidence level.  The best fit value of the $\chi^2$ is very close to the one obtained previously including GW data, implying that the latter is not very important, due to the limited number of data points, and large errors.}
 \label{fig:Pantheon edge-noshell-no GW}
\end{figure}
\end{center}

The peculiar velocity fits in the previous sessions were based on combining luminosity distance observations of GW sources and SNe.
In this session we report the results of fitting only the SNe data, in order to assess the impact that these data have on the results.
As shown in fig.(\ref{fig:Pantheon edge-noshell-no GW}), the impact of removing GW data on the parameters estimation is negligible. This can be explained by the limited number of data points and large errors, which limits the impact on the $\chi^2$. When more GW data will be available we expect the impact to increase.

\section{Conclusions}
We performed a combined analysis of the  luminosity distance of SNe and GW sources,  using different SNe catalogues.  The results of our analysis show that the impact of the GW data is negligible, due to the limited number of data points and large errors. 
In our analysis we do include the effects of a possible local inhomogeneity, and make different assumptions for the value of $H_0$.
We find that the data can be fitted equally well, and in some cases better, by a model with a small local inhomogeneity and the value of $H_0$ obtained by the Planck mission.

We have first fitted directly the apparent magnitude of SNe, and then introduced a method to reconstruct the peculiar velocity field from the luminosity distance, obtaining similar results with both methods. The results obtained for the direct fit of the apparent magnitude should nevertheless be preferred, since they do not require any error propagation associated to obtaining the luminosity distance from the apparent magnitude. 

We found a significant difference between the size and depth of the inhomogeneity obtained analyzing the Pantheon data or the Union 2.1 data, which could be due to a difference in the sky coverage of the two datasets.
%We find  a strong statistical evidence of a radial peculiar velocity field consistent with the one produced by a local underdensity with a density contrast of about $-0.48 \pm 0.003$, extending up to a redshift of about $0.083\pm 0.005$, in agreement with previous studies using number counts \cite{Keenan:2013mfa}.
The implications of the existence of this local underdensity could be profound, since not taking it into proper account can produce a mis-estimation of all background cosmological parameters obtained under the assumption of large scale homogeneity, and it can explain for example the apparent discrepancy between different measurements of the Hubble constant \cite{Romano:2016utn}. Most of the GW sources are located outside this inhomogeneity, and are not affected by it. %explaining why the $H_0$ value estimated from these  objects is in agreement with the estimation based on other large scale observations, such as the CMB. 
This is in accord with the theoretical prediction of the effects of a local inhomogeneity, since the leading monopole perturbative effect is proportional to the volume averaged density contrast. The latter  is inversely proportional to the cube of the distance from the point respect to which the monopole is computed, implying that the high redshift luminosity distance is not affected, including the distance of the last scattering surface from which the $H_0$ is estimated with CMB observations.

Our results are robust under different choices of the redshift range of the analyzed data, and the use of different methods to fit the data.
The edge of the inhomogeneity is around the depth of the 2M++ catalogue used to compute the peculiar velocity redshift correction applied to the Pantheon data, which can naturally explain why such procedure did not remove the effects of the inhomogeneity obtained in our analysis.

In the future it will be important to confirm our results using other independent observables such as number counts\cite{Keenan:2013mfa}, to include the effects of possible anisotropies, and investigate further the cause of the difference between results obtained analyzing different SNe datasets.
%Due to the fact that the results of the analysis of shell averaged and  single data points are approximately the same, higher mutipoles are not expected to play a dominant role, but future GW and SN observations  could be used to further investigate the local structure, and in particular to constrain higher multipoles, whose presence is supported by other analysis \cite{McClure:2007vv,Bengaly:2015nwa,Romano:2014iea}. 

\appendix

\section{Peculiar velocity redshift correction and SNe calibration}
We can only measure the apparent magnitude of SNe, and in order to determine the corresponding luminosity distance, required to fit any cosmological model, a calibration is needed, since we do not have a direct measurement of the absolute luminosity.
For low redshift SNe this is achieved by using the period-luminosity relation for Cepheids in the same galaxy of the SNe, which is calibrated using the NGC4258 megamaser angular diameter distance \cite{Reid:2019tiq}. The measurement of the NGC4258 megamaser distance is of fundamental importance for the calibration of the Cepheid-SN Ia distance ladder. If the local Universe were not homogeneous the observed angular diameter distance would be affected, due to the gravitational effects which can be modeled quite precisely \cite{Romano:2016utn} using eq.(\ref{Ddz}), including that of NGC4258.
The angular diameter distance is modified with respect to the background, due to the peculiar motion of the source with respect to the cosmic flow. 
The effects of peculiar velocities can be removed by applying a redshift correction (RC), which corresponds to eliminate the peculiar contribution from  the observed redshift, to obtain a corrected redshift which should just be representative of the Hubble flow, and could be used for background cosmological parameters estimation.

Nevertheless, if the local inhomogeneity extends beyond the depth of the galaxy catalogue used to obtain the peculiar velocity field used for RC, its effects cannot be removed, and any background cosmological parameter obtained assuming homogeneity, because all inhomogeneities effects have supposedly been removed by RC, will be mis-estimated, including the Hubble parameter.
A similar phenomenon can happen for the cosmological constant for example, leading to a correction of the parameters estimated assuming homogeneity \cite{Romano:2011mx}, and a similar correction to the angular diameter distance of NGC4258 \cite{Reid:2019tiq}, or any other anchor used for calibration, could explain the discrepancy with the Planck estimation of $H_0$.
Note that as explained in detail in \cite{Romano:2016utn}, the effects of a local inhomogeneity on the high redshift luminosity distance are negligible, since the volume average of the density contrast tends to zero at infinity, making this effects only important for low redshift observations.

%In order to obtain the distance modulus $\mu=m-M$ from the observed apparent magnitude $m$, it is necessary to know the absolute magnitude $M$, and this is achieved by the above mentioned calibration, which relies crucially on the NGC4258 angular diameter distance.
%If the observed angular  diameter distance requires a correction, also $M$ will change, and consequently $\mu$, and the estimation of $H_0$. 

This correction can be well approximated as \cite{Romano:2016utn}
\be
\frac{\Delta H^{app}(z)}{H_0^{true}}=-\frac{1}{3}f \overline{\delta}(z) \,,
\ee
where $\overline{\delta}$ is the volume average of the density contrast, and shows that an underdensity increases the local apparent  value  with respect to the true background value.
The implication of the above considerations is that when analyzing SNe data with models assuming different values of $H_0$, the absolute magnitude $M$ has to be changed accordingly, as we will discuss in details in the next appendix.

\section{The relation between $H_0$ and $M$}
In order to clarify the relation between the observed quantity $m^{obs}$, and the parameters $M$ and $H$, let's consider the relation between the distance modulus and the luminosity distance
\bea
\log_{10}(D_L) &=& 1 + \frac{\mu}{5} \,, \\
\mu&=&m-M \,,
\eea
and after defining $D_L=d/H_0$ we can obtain 
\bea
m^{obs}&=&(5 \log{d}-5)+(M-5\log{H_0})=f(\Omega_i)+g(M,H_0)-5 \,, \\
f(\Omega_i)&=&5 \log{d} \,,  \\
g(M,H_0)&=&M-5\log{H_0} \label{fg} \,,
\eea
where the  function $f$ depends only on the cosmological parameters $\Omega_i$.
For example for a flat Universe we have
\bea
h(z)&=& \left[\Omega_m(1+z)^3+\Omega_{\lambda}\right]^{1/2} \,,\\
d(z)&=& (1+z)\int^z \frac{dz'}{h(z')} \,,\\
D_L(z)&=&\frac{1}{H_0}d(z) \,.
\eea
For other sets of cosmological parameters $\Omega_i$ numerical integration is required, but the general structure of eq.(\ref{fg}) would still be valid, i.e. it is always possible to write $m^{obs}$ as the sum of two independent functions depending respectively on $\Omega_i$ and $\{H_0,M\}$.
From eq.(\ref{fg}) it is evident that there is a degeneracy between the parameters $H_0$ and $M$, since keeping $f$ constant, different combinations of $\{H_0,M\}$ can explain the  same observational data $m^{obs}$, as long as $g=const$.
In general  the parameters $\{\Omega_i,H_0,M\}$ are independent, so the same observational data $m^{obs}$ could be explained by different sets of values, and a joint analysis is required to obtain the best fit parameters.
Nevertheless at low-redshift the function $f(\Omega_i)$ is only mildly dependent on  $\Omega_i$, since at low-reshift we have that
\bea
D_L(z)&=&\frac{1}{H_0}\left(z+\frac{1-q_0}{2}z^2+ ..\right) \,, \\
q_0&=&\frac{3}{2}\Omega_m-1 \,,
\eea
so that at low redshift we have that $d\approx z$ is approximately independent of $\Omega_i$, which is the Hubble law. This was indeed the reason for the need of high red-shift observations to obtain evidence of dark energy, since only higher order terms in the Taylor expansion of $D_L(z)$  depend on $\Omega_i$, and also the reason why Einstein considered the cosmological constant its biggest mistake, since he did not know about the implications of the observation of higher redshift SNe.

Let's consider now the case of the two models for which the function  $d$ is the same, $d_a=d_b$, where we are denoting with subscripts $a,b$ the values of the parameters for the two models.
As explained above, at low redshit this should always be a good approximation, or it could be just an assumption, as for some of the fits we perform, with $\Omega_m$ fixed.

When the  assumption that $d_a=d_b$ is justified according to the above arguments we get 
\be
g_a=g_b=M_a-5\log_{10}{H_a}=M_b-5\log_{10}{H_b}  \,,
\ee
from which we finally obtain the following useful relations
\bea
    D_L^{a} &=& D_L^{b} \frac{H_a}{H_b} \,, \\ \label{HM}
    \mu_{a} &=& \mu_{b} + 5\log_{10}\left(\frac{H_{b}}{H_{a}}\right) \,, \\
    M_{a} &=& M_{b} + 5\log_{10}\left(\frac{H_{a}}{H_{b}}\right) \,, 
\eea
where the subscripts a and b correspond to different values of the parameters, i.e. $\{D_L^{a},H_{a},M_a\}$ and $\{D_L^{b},H_{b},M_b\}$. These formulae are related to the well known degeneracy \cite{Scolnic:2017caz} between $H_0$ and $M$. 
Note that a similar relation for $M$ was derived in \cite{Benevento:2020fev}, but claimed to be valid only for $\Lambda CDM$ models, while our derivation is completely model independent, and as explained above, we also observe that it is is valid only if $g_a=g_b$, which at high redshift cannot be assumed unless the $\Omega_i$ are fixed for different models. In any case even at high redshift the parameters $\Omega_i$ should not be too different among different models, so it can be considered approximately valid also at high redshift.

%??

For the Pantheon dataset we have taken as reference the set of parameters $\{H_0= 73.24 \pm 1.59,M=-19.25 \pm 0.71\}$ from \cite{Riess:2016jrr},  for Union 2.1 we use $M=-19.32$, and for Planck \cite{Akrami:2018odb} $H_0=67.4\pm 0.5$ .
The values for the different  parameters obtained using the formulae above are given in the Tables \ref{tab:union_cal} and  \ref{tab:pantheon_cal}. 
\begin{table}[H]
    \centering
   \begin{tabular}{lcc}
   \hline
\toprule
Dataset &$ H_0 (km \, s^{-1} Mpc^{-1})$ & M  \\
\midrule
\hline
Riess & \underline{ $73.24 \pm 1.59$ } & \underline{$-19.25 \pm 0.71$}\\
Planck & \underline{ $67.4 \pm 0.5$} & $-19.4 \pm 0.65$  \\
\bottomrule
\hline
\label{Tpar}
\end{tabular}
    \caption{Values of the parameters used in the analysis of the Pantheon data, obtained using the values in \cite{Riess:2016jrr} as reference. The first row shows the values estimated by \cite{Riess:2016jrr}, and the second line the value of $H_0$ estimated by \cite{Akrami:2018odb} and the corresponding value of $M$ obtained from eq.(\ref{HM}). The underlined values are the  publicly available ones and the not underlined are the values inferred using eq.(\ref{HM}).}
    \label{tab:pantheon_cal}
\end{table}

\begin{table}[H]
    \centering
   \begin{tabular}{lcc}
   \hline
\toprule
Dataset &
$H_0 (km \, s ^{-1} \,  Mpc ^{-1}$) &M  \\
\midrule
\hline
Union 2.1 & \underline{ $70$ } &  \underline{ $-19.32$ } \\
Riess & \underline{$73.24 \pm 1.59$} & $-19.2 \pm 0.11$\\
Planck & \underline{$67.4 \pm 0.5$} & $-19.4 \pm 0.04$\\
\bottomrule
\hline
\end{tabular}
    \caption{Values of the parameters used in the analysis of the Union 2.1 data, obtained using the latter as reference. The underlined values are the  publicly available ones and the not underlined are the values inferred using eq.(\ref{HM}).}
    \label{tab:union_cal}
\end{table}

When fitting data with models with different values of $H_0$ we changed the data according to the above equations.
In the case of Pantheon the publicly available data gives $m$, from which $\mu$ is obtained using the corresponding $M$, while for Union 2.1 the distance modulus $\mu$ is given, which can be calibrated according to the above equations as well.
For example, we can use the Pantheon dataset and fit the data assuming a value of $H_0$ equal to the Planck estimation, or use the Union data, and assume a value of $H_0$ equal to the one obtained in \cite{Riess:2016jrr}.
Note that this kind of calibration is not always correctly performed in the literature, leading to a an \emph{implicit bias in the data analysis}. For example analyzing data allowing for a varying $H_0$, without consistently changing $M$ is an inconsistent approach \cite{Kenworthy:2019qwq}.
The above relations are based on the assumption that $d_a=d_b$, so they can be applied to homogeneous models using the same parameters $\Omega_i$, or to the homogeneous regions of a locally inhomogeneous models, as long as in the homogeneous regions the condition $d_a=d_b$ is satisfied, as in the case of a locally inhomogeneous model we consider. Inside the inhomogeneous region $d$ can be different, even at low redshift and if  the parameters $\Omega_i$ are the same, so a full fit of $M$ could be required, but if most SNe are located outside the inhomogeneity, the formulae above should still give a good approximation. 

It should be noted that the Pantheon dataset has been updated to correct an error in the calculation of the redshift correction \cite{Rameez:2019nrd,github}, which was applied to objects far beyond the depth of the galaxy catalogue used to obtain the peculiar velocities, and we use the second corrected version.

\section{Derivation of the formula for $\overline{\delta}(z)$}

In the case of a step density contrast profile of the type given in eq.(\ref{deltastep}), using the low redshift approximation $\chi\approx z/(a H_0)=z(1+z)/H_0$,  the volume average of the density contrast outside the inhomogeneity at a comoving distance $\chi(z)$, can be computed as
\be
\overline{\delta}(\chi)=\frac{3}{4 \pi \chi^3}\int^{\chi}_0 4 \pi \chi'^2 \delta(\chi') d\chi' =\frac{ H_0^3}{[z(1+z)]^3} \delta_{v}\chi_v^3= \delta_{v} \left[ \frac{z_v(1+z_v)}{z(1+z)} \right]^3  \,.
\ee
At low redshift, which is the regime of validity of the approximation, the factor $(1+z_v)/(1+z)$ can be safely neglected as shown in fig.(\ref{deltaz}).

\section{Results of the fits of $m$ using  the covariance matrix}
The results of the fit of the apparent magnitude using the covariance matrix to compute the $\chi^2$ are given in the table below, showing that  the results are approximately the same as the ones obtained ignoring the  covariance,  not affecting significantly the final conclusions.
\begin{table}[H]
\centering
\resizebox{17cm}{!}{
\begin{tabular}{lccc c ccc c c}
\hline
 & \multicolumn{3}{c}{$z_{sup}=0.11$} && \multicolumn{3}{c}{$z_{sup}=0.5$} && \multicolumn{1}{c}{$z_{sup}=1.5$} \\
\cline{2-4} \cline{6-8} \cline{10-10}
& $\delta_v$   & $z_{v}$  & $\chi^2$  & &  $\delta_v$ & $z_{v}$ & $\chi^2$  & &  $\chi^2$    \\
\cline{2-4} \cline{6-8} \cline{10-10}

$m^{\textup{Inh}}(H_0^P)$ & $-0.142 \pm 0.043 $ & $0.056 \pm 0.0002 $ & 222.724 & & $-0.173 \pm 0.040 $ & $0.056 \pm 0.0004$ & 860.382  & &  1031.940 \\
$m^{\textup{Hom}}(H_0^P)$ & -                   & -                   & 233.139 & & -                   & -                  & 878.889  & &  1058.510 \\
$m^{\textup{Hom}}(H_0^R)$ & -                   & -                   & 227.943 & & -                   & -                  & 857.919  & &  1027.149 \\

\hline
\end{tabular}
               }
    \caption{Results of the fit of the Pantheon data including the covariance matrix for $m$. }
    \label{tab: fit m + cov + inhP + homR + 0.11,0.5,1.5 }
\end{table}

\section{Fits of Pantheon data without peculiar velocity redshift correction}
A peculiar velocity redshift correction is applied to the Pantheon data, consisting in subtracting the peculiar velocities inferred from the 2M++ galaxy catalog, while the Union 2.1 is not redshift corrected.
In order to determine if this could be the cause of the difference in the results obtained for the two datasets, we remove the redshift correction using the publicly available data of the peculiar velocity \cite{peculiar_v}.

The table below shows that the results for non redshift corrected data are approximately the same as those for redshift corrected data, excluding that redshift correction could be the cause of the above mentioned data fit results difference.

\begin{table}[H]
\centering
\resizebox{17cm}{!}{
\begin{tabular}{lccc c ccc c c}
\hline
 & \multicolumn{3}{c}{$z_{sup}=0.11$} && \multicolumn{3}{c}{$z_{sup}=0.5$} && \multicolumn{1}{c}{$z_{sup}=1.5$} \\
\cline{2-4} \cline{6-8} \cline{10-10}
& $\delta_v$   & $z_{v}$  & $\chi^2$  & &  $\delta_v$ & $z_{v}$ & $\chi^2$  & &  $\chi^2$    \\
\cline{2-4} \cline{6-8} \cline{10-10}

$m^{\textup{Inh}}(H_0^P)$ & $-0.156 \pm 0.026 $ & $0.056 \pm 1.5\times 10^{-4} $  & 223.968 & & $-0.159 \pm 0.026 $ & $0.056\pm 8.22\times 10^{-4}$ & 867.421  & &  1036.039 \\
$m^{\textup{Hom}}(H_0^P)$ & -                   & -                               & 260.025 & & -                   & -                             & 905.185  & &  1073.810 \\
$m^{\textup{Hom}}(H_0^R)$ & -                   & -                               & 236.855 & & -                   & -                             & 869.541  & &  1040.891 \\
\hline
\end{tabular}
               }
    \caption{Results of the analysis of  the Pantheon data for non redshift corrected data. %In the tab.(\ref{tab: fit m + no cov + inhP + homR + 0.11,0.5,1.5 }) we can see the results for the redshift corrected data.
    }
    \label{tab: fit m + no cov + no zcorr + inhP + homR + 0.11,0.5,1.5 }
\end{table}

\section{Results of fits varying $\Omega_m$}
We report below the results of the fit of the apparent magnitude varying also the parameter $\Omega_m$.

\begin{center}
\begin{figure}[H]
 \centering
 \includegraphics[scale=0.5]{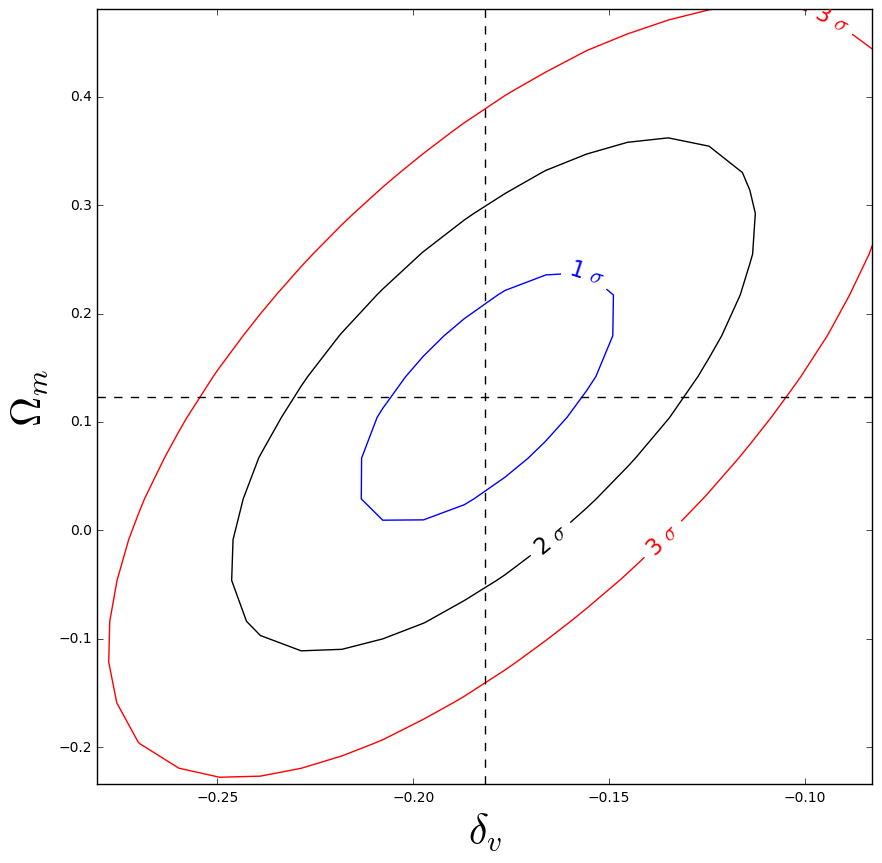}
 \caption[]{Contour plot of $\Omega_m$ and $\delta_v$ for the Pantheon dataset, corresponding to the model $m^{\textup{Inh}}(H_0^{P})$, with $z_{sup}=0.11$.}
 \label{fig:Pantheon + fit m + inhP + Om + 0.11 + contour}
\end{figure}
\end{center}

\begin{table}[H]
\centering

\begin{tabular}{lccccc}
\toprule
\hline
{} &   $\Omega_{m}$ &        $ \delta_{v}$ &      $z_v$ & $\chi^2$  \\
\midrule
\hline
$m^{\textup{Inh}}(H_0^{P})$ &                         $0.123 \pm 0.119$                & -0.182$\pm$0.033 & 0.065$\pm$0.001 &  222.694  \\
$m^{\textup{Hom}}(H_0^{P})$ &                         $0.123 $                &                - &                 - &  276.195  \\
$m^{\textup{Hom}}(H_0^{R})$ &                         $1.45\times 10^{-5} \pm 0.944$   &                - &                 - &  247.288  \\
\hline
\bottomrule
\end{tabular}

    \caption{Results of the fit of the apparent magnitude of the Pantheon dataset with $z_{sup}=0.11$,  without fixing the parameter $\Omega_m$.}
    \label{tab: fit m + Om + inhP + homR + 0.11}
\end{table}

Comparing with Table \ref{tab: fit m + no cov + inhP + homR + 0.11,0.5,1.5 } we can conclude that including $\Omega_m$ in the set of fitted parameters does not affect substantially the main results obtained fixing it, it just gives some slightly different estimation of the parameters of the inhomogeneity. Note that low redshift data is expected not to constrain strongly $\Omega_M$, as we have obtained, since the main evidence of dark energy, and the consequent effect on the value of $\Omega_m=1-\Omega_{\Lambda}$, comes from high redshift SNe.

%tables..........

%%%%%%%%%%%%%%%%%%%%%%%%%%%%%%%%%%%%%%%%%%%%%%%%%%%%%%%%%%%%%%%%%%%%%%%%%%%%%%%%%%%%%%%%%%%%%%%%%%%%%%%%%%%
%%%%%%%%%%%%%%%%%%%%%%%%%%%%%%%%%%%%%%%%%%%%%%%%%%%%%%%%%%%%%%%%%%%%%%%%%%%%%%%%%%%%%%%%%%%%%%%%%%%%%%%%%%%
%\section{Bibliography}

%%%%%%%%%%%%%%%%%%%%%%%%%%%%%%%%%%%%%%%%%%%%%%%%%%%%%%%%%%%%%%%%%%%%%%%%%%%%%%%%%%%%%%%%%%%%%%%%%%%%%%%%%%%
%%%%%%%%%%%%%%%%%%%%%%%%%%%%%%%%%%%%%%%%%%%%%%%%%%%%%%%%%%%%%%%%%%%%%%%%%%%%%%%%%%%%%%%%%%%%%%%%%%%%%%%%%%%
%%%%%%%%%%%%%%%%%%%%%%%%%%%%%%%%%%%%%%%%%%%%%%%%%%%%%%%%%%%%%%%%%%%%%%%%%%%%%%%%%%%%%%%%%%%%%%%%%%%%%%%%%%%
\medskip

\bibliographystyle{h-physrev4}%apsrev4-1long
\bibliography{Bibliography} %%,soko,snova,nu-rev06,parke-ref} 

\end{document}